\documentstyle[psfig]{l-aa}

\def\aeta{A\&A }

\def\apj{ApJ }

\def\aj{AJ }
\def\mn{MNRAS }

\def\apjl{ApJL \rm}

\def\gsim{\lower.4ex\hbox{$\;\buildrel >\over{\scriptstyle\sim}\;$}}
\def\lsim{\lower.4ex\hbox{$\;\buildrel <\over{\scriptstyle\sim}\;$}}

\begin{document}
%
%
\thesaurus{11.09.4;11.17.1;11.17.4 Q0439-433;11.08.1}
\title{
A damped Ly$\alpha$ candidate at $z$~$\sim$~0.1 toward Q~0439--433
\thanks{Based on observations
collected at the European Southern Observatory, La Silla, Chile}
\thanks{Based on observations obtained with the NASA/ESA {\sl Hubble Space
Telescope} by the Space Telescope Institute, which is operated
by AURA, Inc., under NASA contract NAS 5-26555}}
\author{Patrick Petitjean\inst{1,2} \and Bertrand Th\'eodore$^{3}$ 
\and Alain Smette$^{4}$ \and Yannick Lespine$^{1}$}
\institute{$^1$Institut d'Astrophysique de Paris - CNRS, 98bis Boulevard 
Arago, F-75014 Paris, France\\
$^2$UA CNRS 173- DAEC, Observatoire de Paris-Meudon, F-92195 Meudon
Principal Cedex, France \\
$^3$ Servide d'A\'eronomie du CNRS,  BP 3, F-91371 Verri\`eres-le-Buisson 
Cedex, France \\
$^4$ Kapteyn Astronomical Institute, Postbus 800, NL-9700 AV Groningen, The Netherlands
}
\date{ }
\offprints{P. Petitjean}
\maketitle
\markboth{}{}
\begin{abstract}
  We report on the detection of a $z_{\rm gal}$~=~0.101 galaxy
  projected on the sky at 4.2~arcsec (or 5.2$h^{-1}_{100}$~kpc for
  $q_{\rm o}$~=~0.5) from the quasar Q~0439--433 ($z_{\rm
    em}$~=~0.594).  The {\sl HST} spectrum of the quasar shows strong
  Mg~{\sc ii}, Fe~{\sc ii}, Si~{\sc ii}, Al~{\sc ii} and C~{\sc iv}
  absorption lines at the same redshift as the galaxy. The equivalent
  width ratios of the low ionization lines indicate that this system
  is probably damped with a neutral hydrogen column density of
  $N$(H~{\sc i})~$\sim$~10$^{20}$~cm$^{-2}$. The C~{\sc iv} doublet
  presents a complex structure, and in particular a satellite with a
  velocity $v$~=~1100~km~s$^{-1}$ relative to the galaxy. Additional
  {\sl HST} and redshifted 21cm observations of this QSO-galaxy pair
  would offer an ideal opportunity to study the morphology of a damped
  absorber and the kinematics of the halo of a low-redshift 
  galaxy.  
\end{abstract}


\section{Introduction} \label{intr}
QSO absorption line systems probe the baryonic matter over most of the
history of the Universe (0~$<$~$z$~$\la$~5).  The so-called damped
Ly$\alpha$ (hereafter DLA) systems are characterized by a very large
H~{\sc i} column density ($N$(H~{\sc
  i})~$\ga$~2$\times$10$^{20}$~cm$^{-2}$), similar to the one usually
seen in local spiral disks. Such systems at large redshift
($z$~$\sim$~2--3) are thought to be produced by proto-galactic
disks.  The main argument in favor of this conclusion is that the
cosmological density of gas associated with these systems is of the
same order of magnitude as the cosmological density of stars at
present epochs (Wolfe 1996).  Moreover they present a metallicity $Z
\sim 1/10 ~ Z_\odot$ (Pettini et al.  1994), while strong metal line
systems have been demonstrated to be associated with galaxies at low
and intermediate $z$ (e.g.  Bergeron \& Boiss\'e 1991). The
detailed study of low-$z$ DLA systems is thus crucial as
information can be gathered both on the absorbing gas via optical and
UV spectroscopy (kinematics, ionization state) and the associated
galaxy (impact parameter, morphology, star formation activity).
However very few DLA systems are known at low-$z$ (Lanzetta et
al. 1995, Rao et al. 1995).  Moreover identifying the galaxy
associated with low-$z$ DLA systems is not easy as it often lies
very close to the quasar (Steidel et al.
1994). Indeed, using {\sl HST} high spatial resolution images of the
field of seven quasars whose spectra present DLA lines at
intermediate redshifts (0.4~$\la$~$z$~$\la$~1),
Le~Brun et al. (1996) show that, in all cases, at least one galaxy
candidate is present within 4~arcsec from the quasar, the closest
being at $\sim$0.75~arcsec. There is no dominant morphological type in
their sample: three
candidates are spiral galaxies, three are compact objects and two are
amorphous low surface brightness galaxies.
\par\noindent
Here, we report on the detection of a galaxy projected at 4.2~arcsec (or
5.2~$h^{-1}_{100}$~kpc for $q_{\rm o}$~=~0.5 at $z$~=~0.101) from the
quasar Q~0459--433. The galaxy has the same redshift as a
strong Fe~{\sc ii}-Mg~{\sc ii}-C~{\sc iv} metal line system detected
in {\sl HST-FOS} spectra.
\begin{figure}
 \centerline{\vbox{
 \psfig{figure=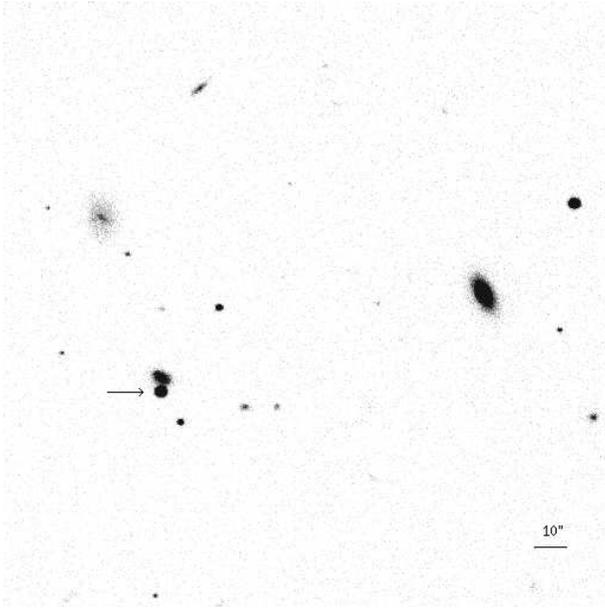,height=9.cm,angle=0}
 }}
\caption[]{B-band image taken with EMMI on the NTT. North is at the top,
East on the left. The quasar is marked.
The scale is given by a tick mark of size 10~arcsec.}
\end{figure}
\section{Observations} \label{s2}
B, V, R and I broad-band images of the rich field around Q~0439--433
and spectra of the QSO and galaxy were obtained on 20 and 22 February
1996, with EMMI mounted on the NTT. Fig.~1 shows the B-band image
after flat-fielding and flux-calibration. The scale is given by a tick
mark of size 10~arcsec.  The quasar is marked with an arrow; a galaxy
is seen 4.2~arcsec north of it. B, V, R and I magnitudes for the
quasar and the galaxy are 17.21, 16.84, 16.75, 16.44 and 18.76, 17.74,
17.24, 16.74 respectively with an error of 0.05 magnitude. The galaxy
luminosity is thus $\sim 0.45 ~ L_*$.  Its inclination measured on the
R and I images is $51 \pm 3^\circ $ assuming this is a spiral galaxy.
\par 
Mg~{\sc ii}$\lambda$2797, [Ne~{\sc v}]$\lambda$3426, [O~{\sc
  ii}]$\lambda$3727 and [Ne~{\sc iii}]$\lambda$3869 emission lines
from the quasar are detected at a mean redshift of $z_{\rm
  em}$~$=$~0.594.  The galaxy spectrum is displayed in Fig.~2. It
presents a spatially resolved emission line with $w_{\rm obs} =
2.8\pm0.3$ \AA\ at $\lambda$5353.08 (on top of a broad absorption
feature), identified as H$\beta$ at $z_{\rm gal}$~=~0.1012; this value
is confirmed by the presence of the Mg~{\sc i} absorption blend and
the Ca~+~Fe E absorption band at 5700.3 and 5800.9~\AA~ for a redshift
of 0.1014 and 0.1009 respectively.
\par 
The infrared observations were carried out on 27-28 January 1995, with
the ESO IRAC2B camera (Moorwood et al. 1992) mounted on the 2.2~m
telescope.  The detector is a 256$\times$256 NICMOS3 array. We used
lens C in the K' and J--bands with a resolution of 0.52~arcsec per
pixel.  The seeing was typically 1~arcsec or slightly better.  Every
2~min, we moved the object on the array by about 10~arcsec.  The image
of the sky has then been obtained by median-filtering all the images.
Each image has then been dark-subtracted and flat-fielded using the
normalized sky-image.  Accurate registration is obtained by
determining the peak of the 2D correlation function between images.
The total integration time is 100~min in K' and 25~min in J. The final
result is displayed in Fig.~3. The magnitudes of the quasar and the
galaxy in J and K' are 16.1, 14.5 and 16.2, 14.7 respectively.
\par
\begin{figure}
\centerline{\vbox{
\psfig{figure=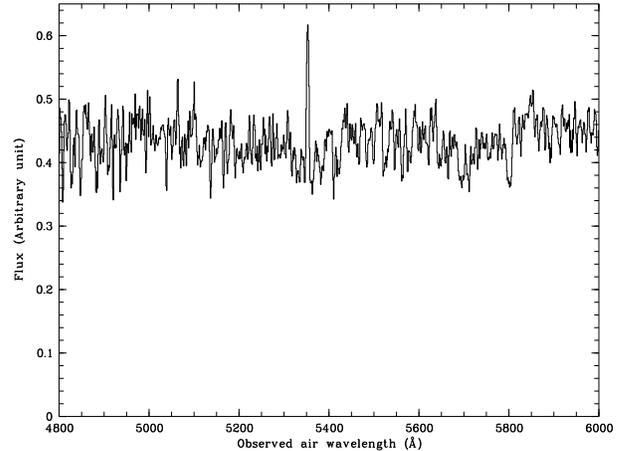,height=6.5cm,width=9.cm,angle=270}
}}
\caption[]{Portion of the spectrum of the galaxy taken with EMMI on the NTT.
  An emission line clearly appears at $\lambda$5353.08 on top of a
  broad absorption feature and is identified as H$\beta$ at $z_{\rm
    abs}$~=~0.1012.}
\end{figure}
Spectroscopic observations of Q~0439--433, yet unpublished,
were accessed from the {\sl HST} archive. 
The observations were made using the Faint Object
Spectrograph  with the G190H grating (over the wavelength range
1600--2300~\AA, for a resolution of 1.35~\AA~ FWHM) and with the G270H
grating (over the wavelength range 2250--3250~\AA, for a resolution of
1.92~\AA~ FWHM). The data were calibrated using the standard pipeline
reduction techniques.  The zero point of the wavelength scale was
determined requiring that Galactic interstellar absorptions occur at
rest. We concentrate on the $z_{\rm abs}$~=~0.101 system. Table~1
gives the characteristics of the lines detected in this system
(observed wavelengths and equivalent widths, errors from the adjacent
noise, identifications and corresponding redshifts). Fig.~4 shows
portions of the spectrum with the lines of interest.

\begin{figure}
\centerline{\hbox{
\psfig{figure=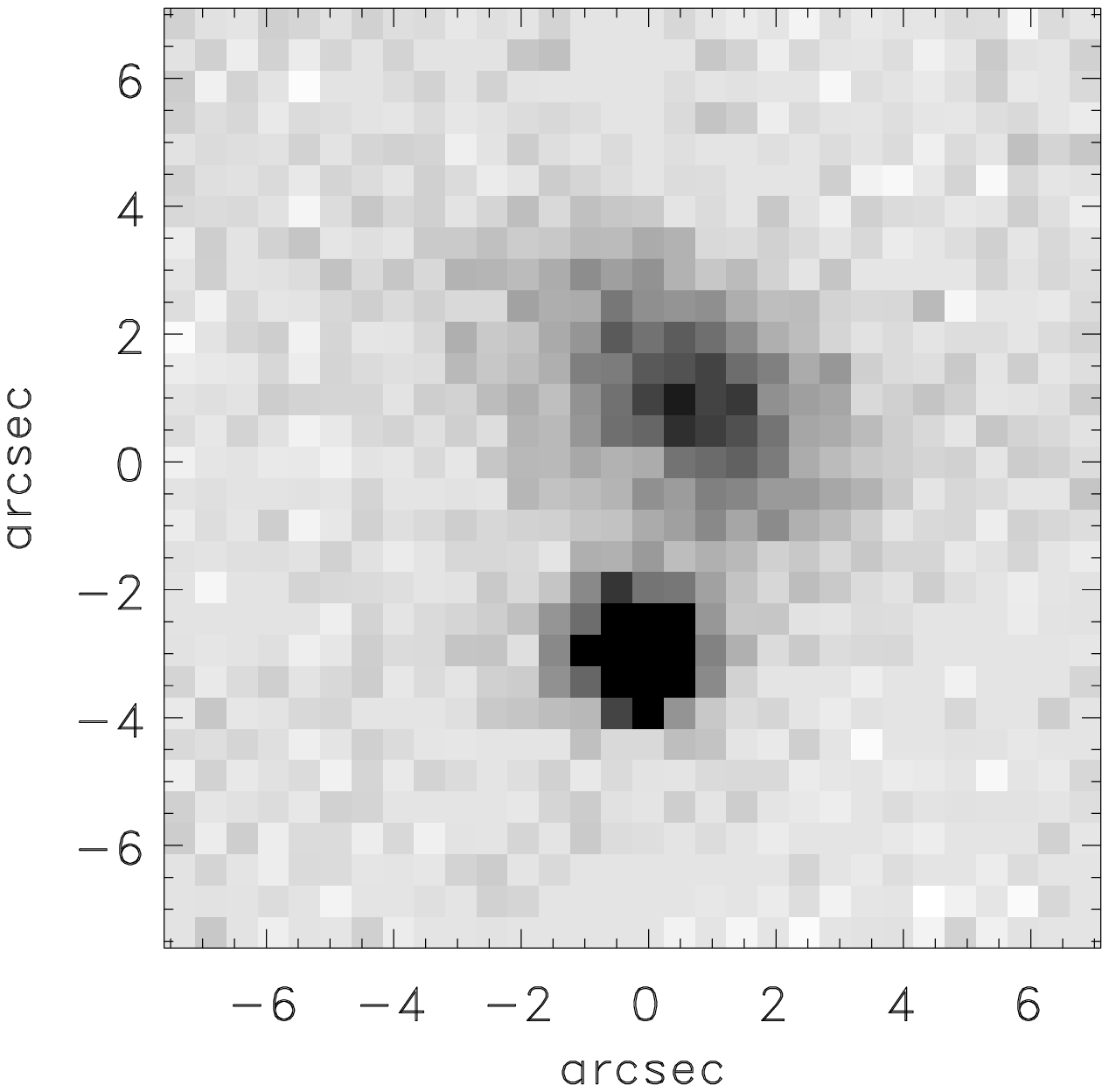,height=4.cm,angle=0}
\psfig{figure=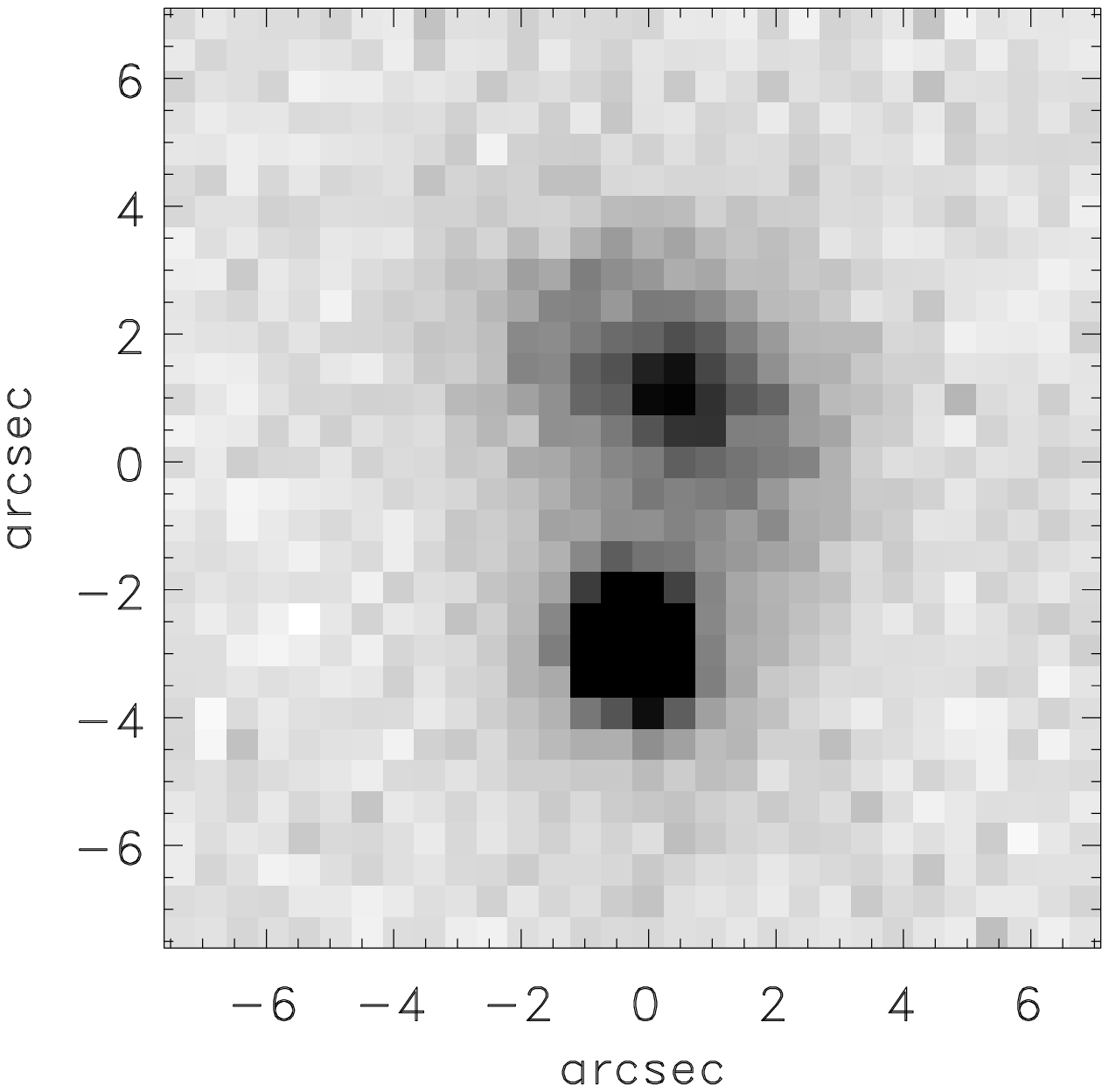,height=4.cm,angle=0}
}}
\caption[]{Images J (left) and K' (right) taken with IRAC2
at the ESO 2.2~m.}
\end{figure}
\section{Results } \label{s3}
\subsection{A galaxy at $z$~=~0.3848}
During the spectroscopic observation, the long-slit was oriented
north-south and happened to intercept a faint object located
25.8~arcsec north of the quasar (see Fig.~1). The spectrum presents an
emission line at $\lambda$5161.3. The identification with [O~{\sc
  ii}]$\lambda$3727 gives a redshift of $z_{\rm gal}$~=~0.3848. There
is no absorption line in the {\sl HST} data at this redshift with a
3$\sigma$ observed equivalent width limit of 0.3~\AA. The projected distance
between the object and the line of sight to the quasar is
324~$h_{100}^{-1}$~kpc ($q_{\rm o}$~=~0.5).  This large distance is
consistent with the idea that the metal line systems at
intermediate redshift arise in galaxy halos with radius of the order of
35~$h_{100}^{-1}$~kpc (e.g. Bergeron \& Boiss\'e 1991, Steidel
1993).\par%
\begin{figure}
\centerline{\vbox{
\psfig{figure=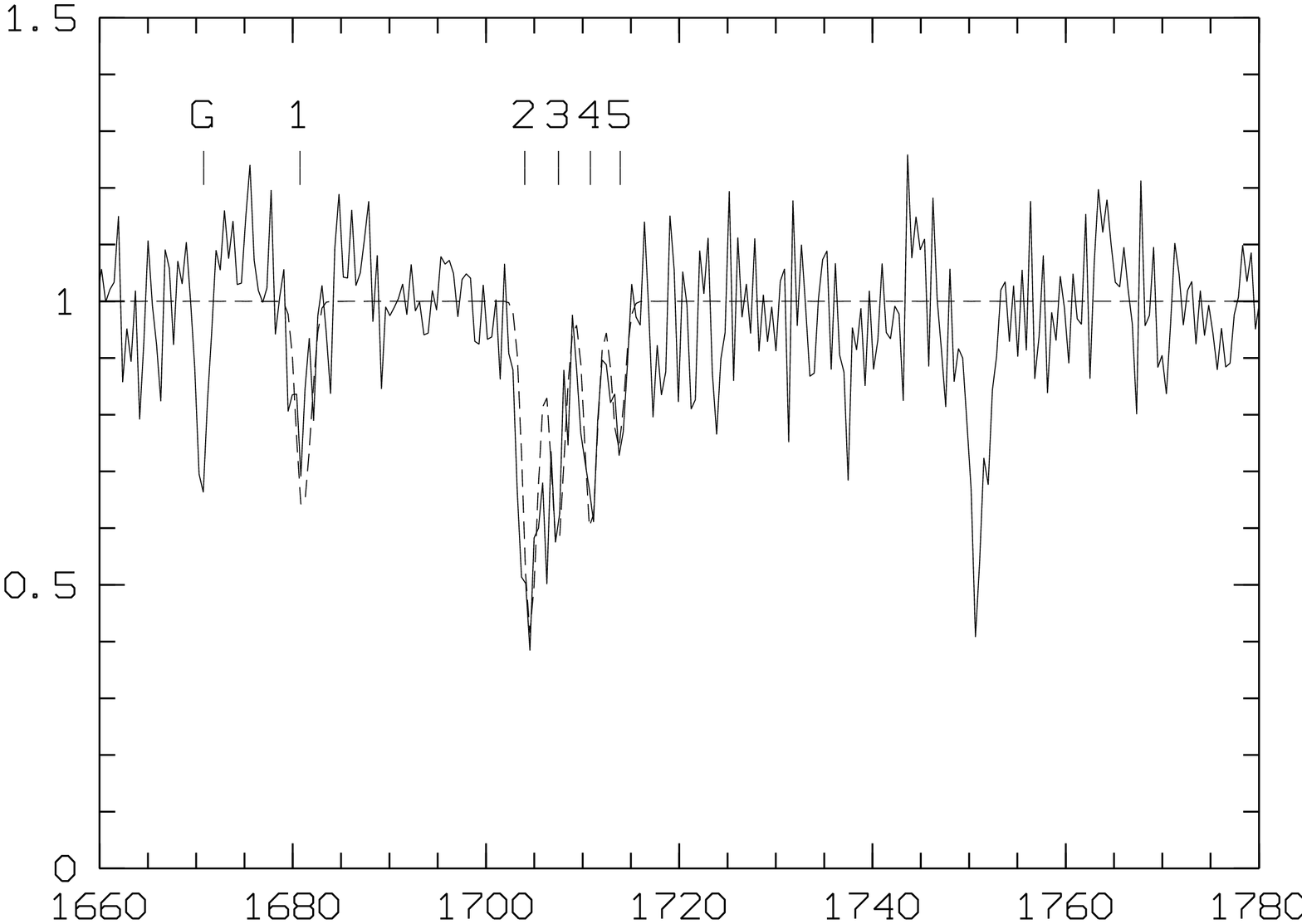,height=3.6cm,width=7.cm,angle=270}
\psfig{figure=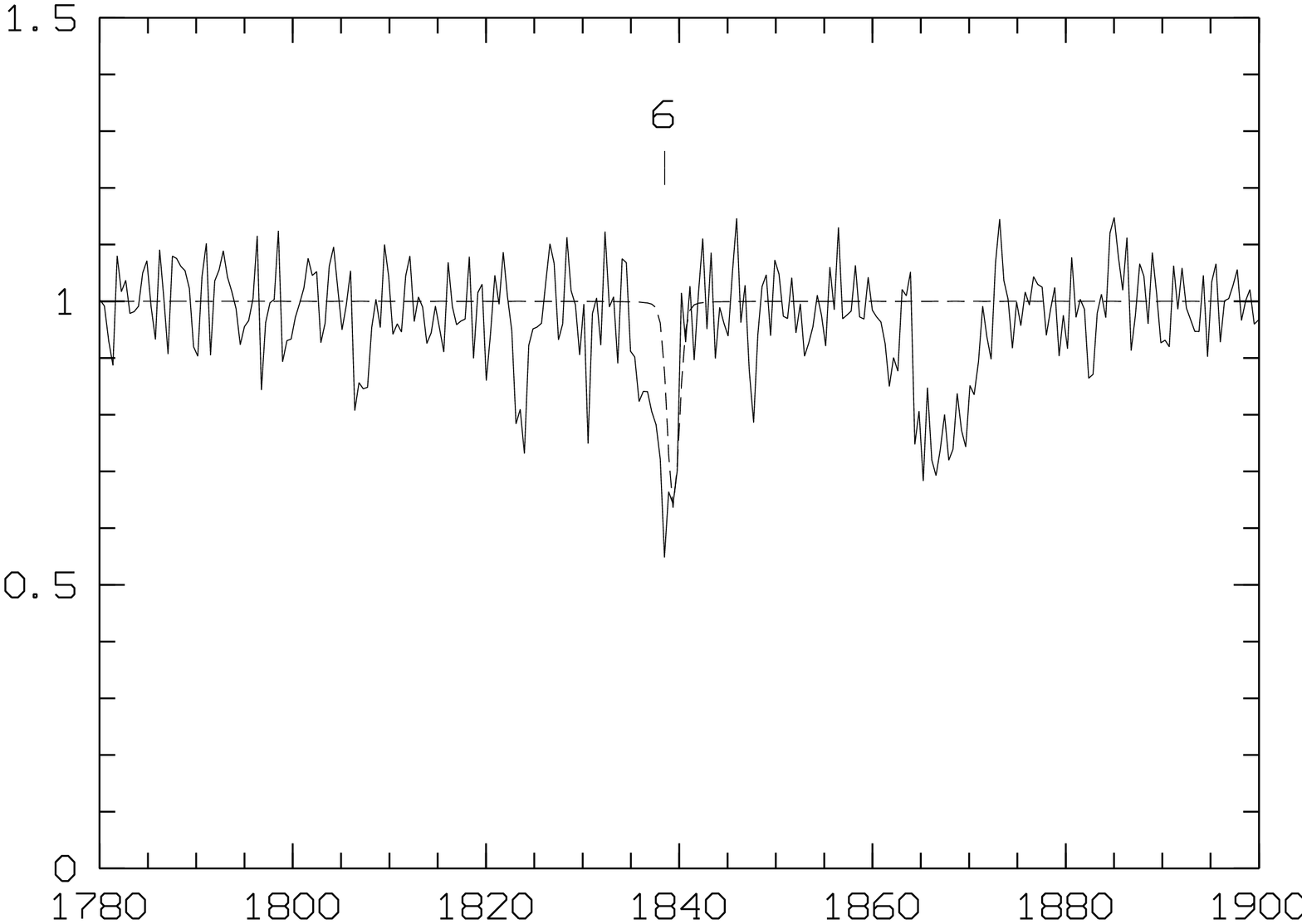,height=3.6cm,width=7.cm,angle=270}
\psfig{figure=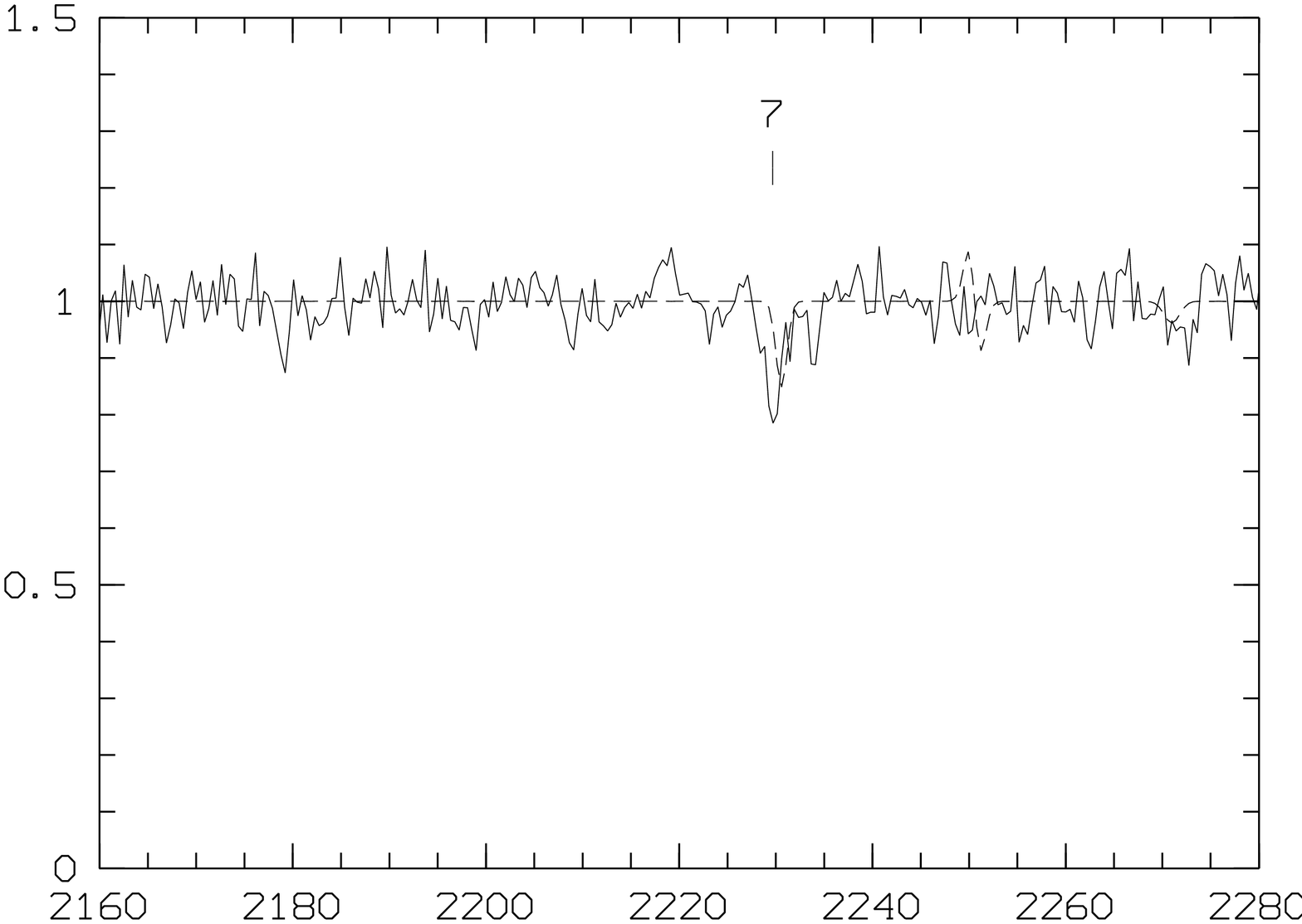,height=3.6cm,width=7.cm,angle=270}
\psfig{figure=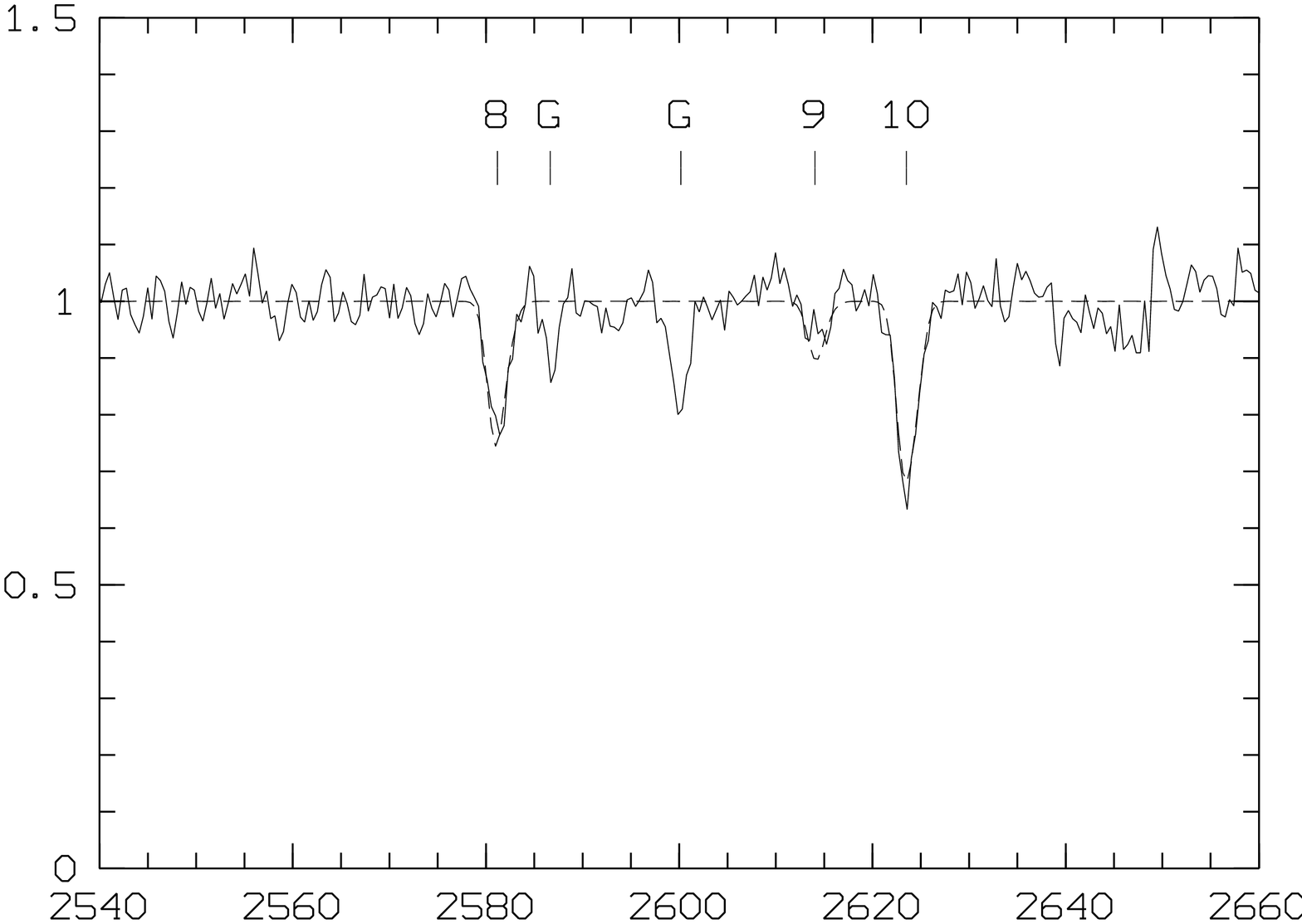,height=3.6cm,width=7.cm,angle=270}
\psfig{figure=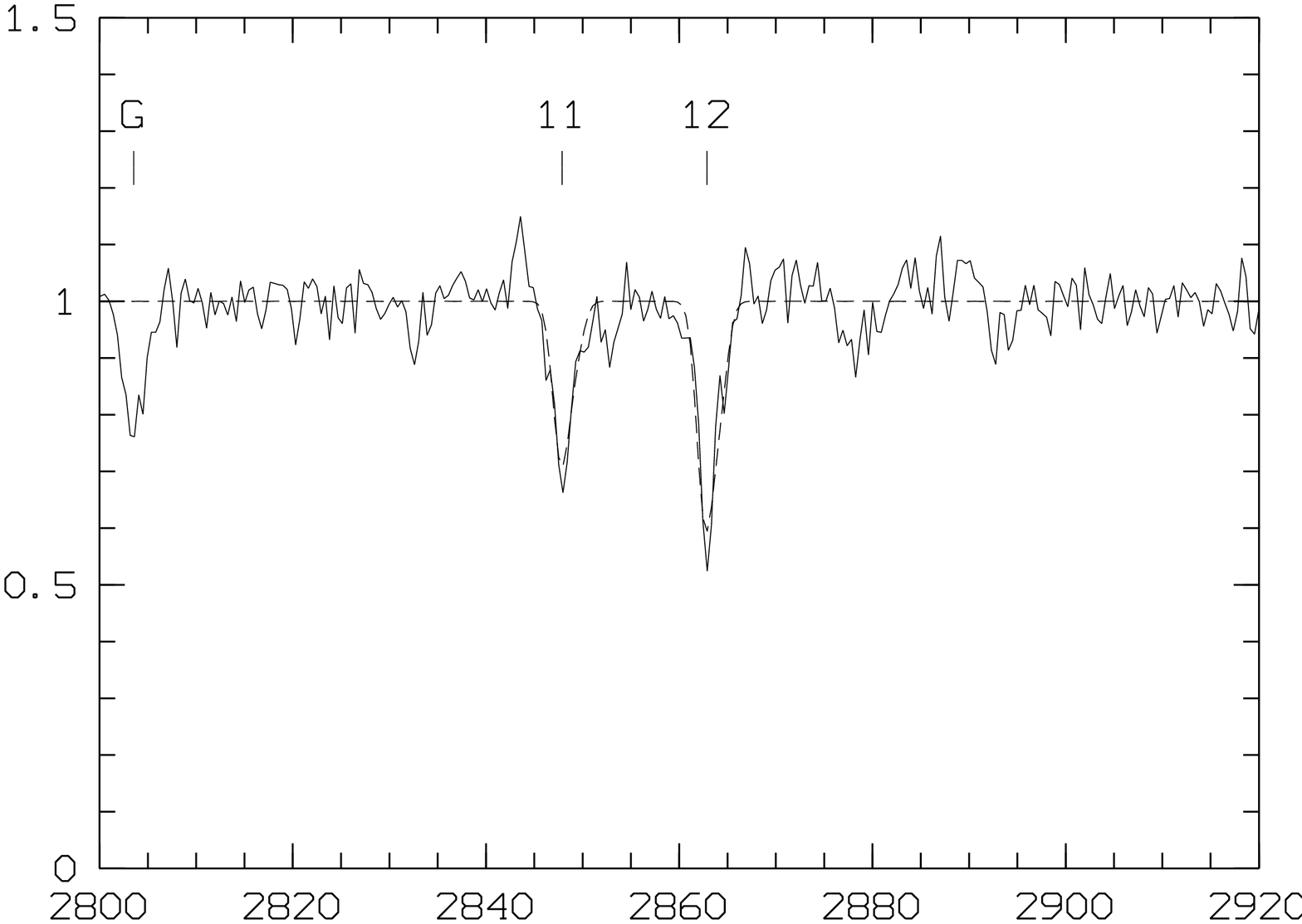,height=3.6cm,width=7.cm,angle=270}
\psfig{figure=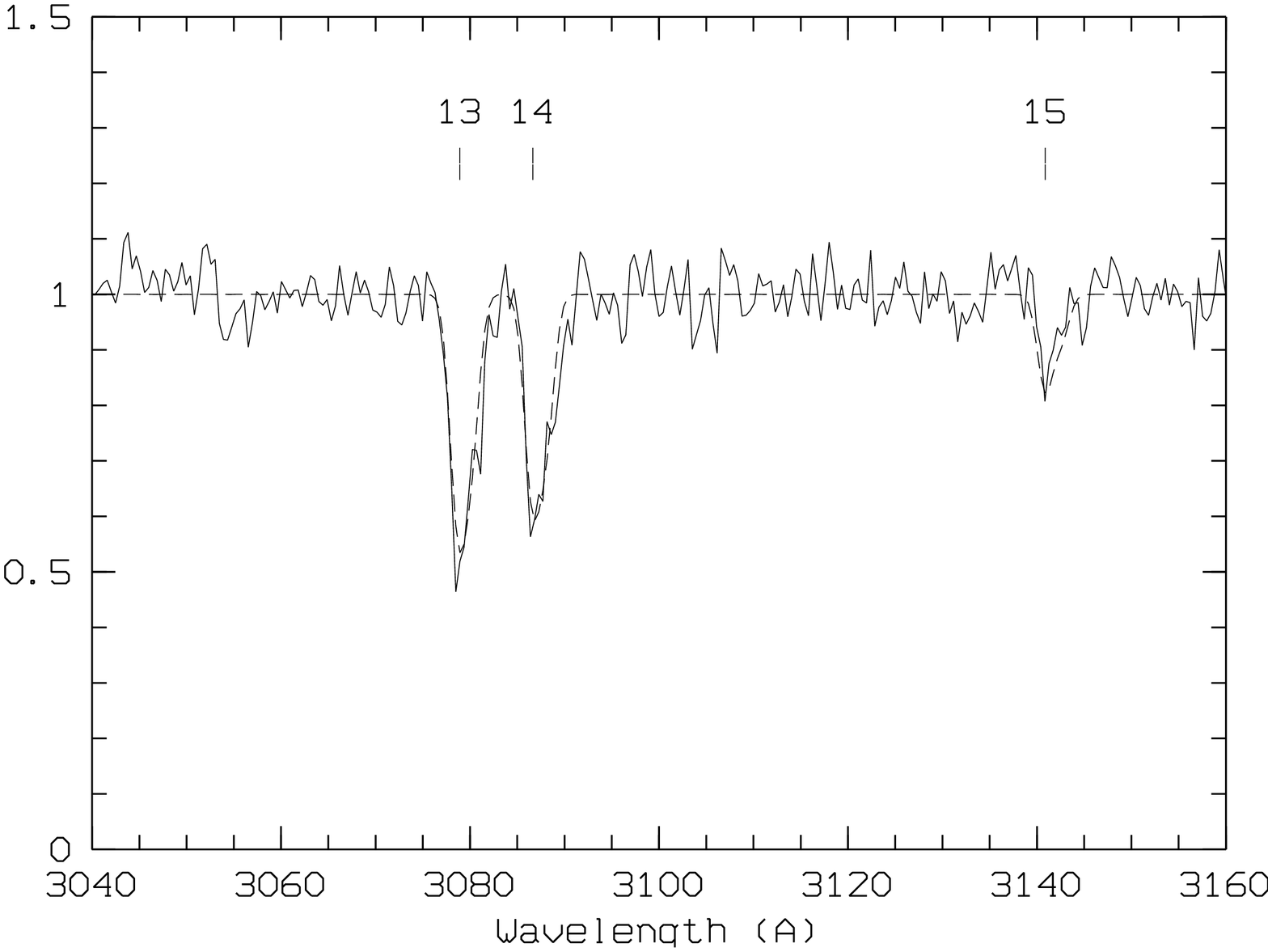,height=3.6cm,width=7.cm,angle=270}
}}
\caption[]{Portions of the G190H and G270H {\sl HST} spectra. 
Lines marked with a G symbol correspond to galactic absorptions. 
Absorption lines from the $z$~$\sim$~0.101 system are numbered as
in Table~1. The dotted line is a fit to the lines described in the text
and with column densities given in Table~2.}
\end{figure}
\begin{table}
\caption[ ]{Absorption lines in the quasar spectrum belonging to the
the $z_{\rm abs}$~=~0.101 system} 
\begin{tabular}{r l r r l r}
\hline
\multicolumn{1}{c}{N.}&
\multicolumn{1}{c}{$\lambda_{\rm obs}$}&\multicolumn{1}{c}{$w_{\rm obs}$}&
\multicolumn{1}{c}{$\sigma_{\rm obs}^a$}&\multicolumn{1}{c}{Identification}&
\multicolumn{1}{c}{$z_{\rm abs}$}\\
\multicolumn{1}{c}{ }&
\multicolumn{1}{c}{\AA}&\multicolumn{1}{c}{\AA}&
\multicolumn{1}{c}{\AA}&\multicolumn{1}{c}{}&
\multicolumn{1}{c}{}{}\\
\hline
\\
1 & 1680.75  & 0.71 & 0.14 & SiII$\lambda$1526 & 0.10090           \\
2 & 1704.91  & 1.40  & 0.14 & CIV$\lambda$1548  & 0.10122 \\
3 & 1707.40  & 0.90  & 0.14 & CIV$\lambda$1550  & 0.10100 \\
4 & 1710.85  & 0.70 & 0.14 & CIV$\lambda$1548  & 0.10506 \\
5 & 1713.70  & 0.50 & 0.14 & CIV$\lambda$1550  & 0.10506 \\
6 & 1838.77  & 1.25 & 0.12 & AlII$\lambda$1670 & $^{\rm b}$ \\
7 & 2229.60: &      &      & ZnII$\lambda$2026 & $^{\rm b}$ \\
8 & 2581.19  & 0.60 & 0.10 & FeII$\lambda$2344 & 0.10109    \\
9 & 2614.05  & 0.22 & 0.10 & FeII$\lambda$2374 & 0.10090    \\
10 & 2623.51  & 0.98 & 0.10 & FeII$\lambda$2382 & 0.10104     \\
11 & 2847.89  & 0.94 & 0.10 & FeII$\lambda$2586 & 0.10099     \\
12 & 2862.87  & 1.07 & 0.10 & FeII$\lambda$2600 & 0.10103     \\
13 & 3078.92  & 1.64 & 0.10 & MgII$\lambda$2796 & 0.10105     \\
14 & 3086.64  & 1.39 & 0.10 & MgII$\lambda$2803 & 0.10098    \\
15 & 3140.89  & 0.29 & 0.10 & MgI$\lambda$2852  & 0.10105    \\
(NTT) & 4333.74: & 0.39 & 0.15 & CaII$\lambda$3934 & 0.10139    \\
\\
\hline
\multicolumn{5}{l}{$^{\rm a}$ 1$\sigma$ detection limit } \\
\multicolumn{5}{l}{$^{\rm b}$ Blended} \\
%
\end{tabular}
\end{table}
\begin{table}
\caption[ ]{Column densities (cm$^{-2}$) in the $z_{\rm abs}$~=~0.101 system} 
\begin{tabular}{r r r r r r r}
\hline
\multicolumn{1}{c}{Velocity$^{\rm a}$}&\multicolumn{1}{c}{C~{\sc iv}}&
\multicolumn{1}{c}{Si~{\sc ii}}&
\multicolumn{1}{c}{Fe~{\sc ii}}&
\multicolumn{1}{c}{Mg~{\sc ii}}&\multicolumn{1}{c}{Mg~{\sc i}}&
\multicolumn{1}{c}{Zn~{\sc ii}}\\
\hline
\\
-16  & 14.7 & 15.5: & 14.7 & 14.8 & 12.6 & 13.0:  \\
136  & 13.8 & 13.8  & 13.6 & 14.3 & 12.0 & $<$12.5 \\
1100 & 14.4 & & & & & \\
\hline
\multicolumn{5}{l}{$^{\rm a}$ relative to $z$~=~0.1010 in km~s$^{-1}$} \\
\end{tabular}
\end{table}
\subsection{The absorption system and the associated galaxy 
at $z_{\rm abs}$~=~0.101}
%
%
Strong Fe~{\sc ii} and Mg~{\sc ii} lines are present in the HST
spectrum with very good redshift agreement. The mean redshift is
$z_{\rm abs}$~=~0.1010.  It is apparent from Fig.~4 that the lines are
blends of several components.  Given the limited resolution of the
spectra, it is only possible to perform a rough decomposition. We
obtain however a good overall fit with two components with Doppler
parameters $b$~=~15~km~s$^{-1}$, separated by 150~km~s$^{-1}$.
The derived column densities are given in Table~2.\par The ratio
$w$(Fe~{\sc ii}$\lambda$2382)/$w$(Mg~{\sc ii}$\lambda$2796) $\sim$ 0.6
indicates that the system is certainly of fairly high H~{\sc i} column
density ($N$(H~{\sc i})~$>$~10$^{19}$~cm$^{-2}$, Bergeron \&
Stasi\'nska 1986). The detection of a strong Mg~{\sc i}$\lambda$2852
absorption line (rest equivalent width $w_{\rm r}$~=~0.26~\AA) further
supports this conclusion. Indeed the Fe~{\sc ii}, Mg~{\sc ii} and
Mg~{\sc i} equivalent widths are very similar to those observed in the
21~cm absorber ($N$(H~{\sc i})~$\sim ~ 3 \times 10^{20}$~cm$^{-2}$) at
$z_{\rm abs}$~=~0.395 towards Q~1229$-$021 (Lanzetta \& Bowen 1992).
There is an absorption feature at $\lambda$2229.6 which we would normally
not attribute to Zn~{\sc ii}$\lambda$2026 at $z$~=~0.101 because the
wavelength discrepancy (120~km~s$^{-1}$) is slighly larger than the
resolution.  However the feature has a red wing that could be due to
Zn~{\sc ii}, in which case we find log~$N$(Zn~{\sc ii})~$\sim$~13.3.
There is an absorption feature at $\lambda$4333.74 in the NTT spectrum
that could be identified with Ca~{\sc ii}$\lambda$3934 at $z_{\rm
  abs}$~=~0.10139.  The associated Ca~{\sc ii}$\lambda$3969 is
below our detection limit of $w_{\rm obs} = 0.15$ \AA\ (1~$\sigma$).  
The non-detection of Ca~{\sc ii} in a strong Mg~{\sc ii}
system would not be surprising however.  Bowen et al. (1991) have
shown that the distribution of the calcium-absorbing gas is quite
inhomogeneous in present-day galaxies even at very small impact
parameters ($\rho < 5 ~ h_{100}^{-1} ~$kpc).
\par 
There is a strong C~{\sc iv} system coincident in redshift with the
low ionization lines. The C~{\sc iv} and Mg~{\sc ii} doublets are
spread over  $\sim$~150~km~s$^{-1}$. However the
velocity distribution of the gas is different in C~{\sc iv} and
Mg~{\sc ii} (Doppler parameters of 60~km~s$^{-1}$ were 
used to fit the C~{\sc iv} lines) as often observed in DLA
systems where Mg~{\sc ii} and  C~{\sc iv} are thought to probe galactic disks
and haloes respectively (e.g. Turnshek et al. 1989).
The low ionization lines have
equivalent width typical of the absorption lines produced by the Milky
Way gas in the spectra of extra-galactic objects (Savage et al. 1993).
However the C~{\sc iv} lines are much stronger in the present system.  
Moreover there is an additional C~{\sc iv} system
redshifted relative to the galaxy by about 1100~km~s$^{-1}$.  Its
origin is unclear but could be due to
a companion. Several candidates are present in the field within
35~arcsec (or 45$h^{-1}_{100}$~kpc, see Fig.~1). High
resolution imaging of the field with {\sl HST} is needed to
investigate the morphology of the galaxy.
Indeed, although a faint spiral feature can be
seen in the K'-band and optical images, the morphology of the galaxy
is unclear. The colours however are indicative of a spiral
galaxy.\par
\section{Conclusion } \label{s4}
It seems likely that the $z_{\rm abs}$~=~0.101 Mg~{\sc ii}-Fe~{\sc
  ii}-C~{\sc iv} absorption system detected in the spectrum of
Q~0439-433 is a damped Ly$\alpha$ system of moderate column density
($N$(H~{\sc i})~$\sim$~10$^{20}$~cm$^{-2}$). This conclusion is
supported by the large $w_{\rm r}$(Fe~{\sc ii})/$w_{\rm r}$(Mg~{\sc
  ii})~$\sim$~0.6 equivalent width ratio, the presence of a strong
Mg~{\sc i} absorption and strengths of the absorptions by low
ionization species similar to what is observed for damped systems. The
absorption in the X-ray spectrum of Q~0439-433 corresponds to
$N$(H\,{\sc i})$ = (2.3\pm0.8) \times 10^{20}$cm$^{-2}$ (Wilkes et al.
1992) with a Galactic contribution of $(1.3 \pm 0.1) \times
10^{20}$cm$^{-2}$ (Lockman \& Savage 1995).  This is consistent with
the above estimate of the H~{\sc i} column density.  Q~0439--433 is a
flat-spectrum radio-source with a flux of 0.3 Jy at 2.7 GHz (Peterson
\& Bolton 1972): it is then possible to carry out a detailed study of
the kinematics through the 21cm absorption line that is sensitive to
gas with low spin temperature; on the other hand, HST+STIS spectra
should provide an accurate determination of the total H\,{\sc i}
column density, and in turn may lead to evaluate the H\,{\sc i} spin
temperature.  This QSO-galaxy pair is an ideal case to
study the morphology of a damped absorber and the kinematics of the halo 
of a low-redshift galaxy.
\begin{acknowledgements}
  AS thanks financial support under grant no.  781-73-058 from ASTRON
  which is funded  by the NWO.
\end{acknowledgements}

\end{document}